\documentclass[%
 reprint,
 amsmath,amssymb,
 aps,
 nofootinbib
]{revtex4-1}

\usepackage{graphicx}% Include figure files
\usepackage{dcolumn}% Align table columns on decimal point
\usepackage{bm}% bold math
\usepackage{braket}
\usepackage{amsmath}
\usepackage{amssymb}
\usepackage{enumerate}
\usepackage{booktabs}
\usepackage{txfonts}
\usepackage[update,prepend]{epstopdf}
\usepackage[english]{babel}
\usepackage{color}
\usepackage{listings}
\usepackage{qcircuit}
\usepackage[subrefformat=parens]{subcaption}
\usepackage{caption}
\usepackage{ulem}
\makeatletter
\def\hlinewd#1{%
	\noalign{\ifnum0=`}\fi\hrule \@height #1 %
	\futurelet\reserved@a\@xhline}
\makeatother

\begin{document}

\preprint{APS/123-QED}

\title{Reduction of Qubits in Quantum Algorithm for Monte Carlo Simulation by Pseudo-random Number Generator}

\author{Koichi Miyamoto}
\email{koichi-miyamoto@fintec.co.jp}
\affiliation{Mizuho-DL Financial Technology Co., Ltd.\\ 2-4-1 Kojimachi, Chiyoda-ku, Tokyo, 102-0083, Japan}

\author{Kenji Shiohara}
\email{k-shiohara@particle.sci.hokudai.ac.jp}
\affiliation{Division of Physics, Graduate School of Science, Hokkaido University\\
 Sapporo, Hokkaido, 060-0810, Japan}

\date{\today}

\begin{abstract}
It is known that quantum computers can speed up Monte Carlo simulation compared to classical counterparts.
There are already some proposals of application of the quantum algorithm to practical problems, including quantitative finance.
In many problems in finance to which Monte Carlo simulation is applied, many random numbers are required to obtain one sample value of the integrand, since those problems are extremely high-dimensional integrations, for example, risk measurement of credit portfolio.
This leads to the situation that the required qubit number is too large in the naive implementation where a quantum register is allocated per random number.
In this paper, we point out that we can reduce qubits keeping quantum speed up if we perform calculation similar to the classical one, that is, estimate the average of integrand values sampled by a pseudo-random number generator (PRNG) implemented on a quantum circuit.
We present not only the overview of the idea but also concrete implementation of PRNG and application to credit risk measurement.
Actually, reduction of qubits is a trade-off against increase of circuit depth.
Therefore full reduction might be impractical, but such a trade-off between speed and memory space will be important in adjustment of calculation setting considering machine specs, if large-scale Monte Carlo simulation by quantum computer is in operation in the future.
\end{abstract}

\pacs{Valid PACS appear here}% PACS, the Physics and Astronomy
                              
\maketitle

\section{\label{sec:intro}Introduction}
Among applications of quantum computers to numerical problems providing higher speed than classical computation is Monte Carlo simulation \cite{Montanaro}.
It has been shown that estimation error in the quantum-based Monte Carlo is proportional to $O(N^{-1})$, where $N$ is the number of computational steps, compared with $O(N^{-1/2})$ in the classical one.
Quantitative finance is one of the fields where Monte Carlo simulation is heavily used and there are some proposals to apply the quantum algorithm to financial problems, for example, risk measurement of portfolio \cite{Woerner,EggerEtAl} and derivative pricing \cite{RebentrostEtAl,Stamatopoulos}.

In the application of the quantum algorithm for Monte Carlo to financial problems, required qubit number might be problematic due to the following two points. First, in the methods proposed in the previous works \cite{Woerner,EggerEtAl,RebentrostEtAl,Stamatopoulos}, a quantum register is allocated to represent a random number, so the required qubit number is proportional to the number of the random numbers required to obtain one sample value of the integrand (in other words, the dimension of the integral).
Second, many of the problems in finance are extremely high-dimensional integrations and require many random numbers.
One of the most prominent examples is risk measurement of credit portfolio \cite{EggerEtAl}.
Credit portfolio consists of many loans or debts and banks suffer losses when obligors default.
Banks monitor such credit risks estimating some {\it risk measures}, for example, expected loss (EL), value-at-risk (VaR), which represents percentile point (say, 99\%) of loss distribution, conditional VaR (CVaR), expectation value of loss conditioned it exceeds the VaR, and so on.
One of popular mathematical models describing probability distribution of loss is the Merton model \cite{Merton} and risk measures under the model are usually estimated by Monte Carlo.
We describe the model in the later section, but an important point is that the required number of random numbers to determine a default pattern of obligors is nearly equal to the number of obligors.
In other words, it is necessary to generate as many random numbers as obligors to obtain a sample value of the integrand, that is, loss.
The number of obligors can be $O(10^6)$ for large portfolios, and so is the required random number.
The qubit number of today's largest quantum computer is $O(10)$, so it will be the far future when machines with so many qubits are realized.
Therefore, it is meaningful to consider the possibility to reduce qubits.

In this paper, we propose a way to reduce qubits.
Although we will explain the detail in the next section, we here describe the outline.
In short, it is {\it classical Monte Carlo on a quantum computer}.
In classical Monte Carlo, we usually generate {\it some sampled patterns} of values of random numbers, but {\it not all patterns}.
More concretely, we generate sequences of pseudo-random number (PRN) using some pseudo-random number generator (PRNG) and use each sequence to obtain one sample value of the integrand.
Finally, we calculate the average of the sample values and consider it as an approximation of the integral.
An important point is that in this way we sequentially generate PRNs and do not require the memory space proportional to the number of the required random numbers.
We can do the same thing on a quantum circuit.
That is, we can sequentially generate pseudo-random bit strings on a quantum register and calculate the integrand into another register.
On a quantum computer, we can parallelly perform such computation and finally obtain the superposition of states in which each of the sampled integrand values is realized on a register.
Then, we can estimate the average of sample values by the quantum amplitude estimation methods \cite{Bassard,Suzuki}, which are commonly used in the quantum algorithm for Monte Carlo.
This procedure leads to the same estimation result as classical Monte Carlo, but with quadratic speedup.

We present not only the idea but also concrete implementation.
We propose an example of PRNG which can be easily implemented on a quantum circuit.
It is {\it permuted congruential generator (PCG)} \cite{PCG} and explained in detail in a later section.
This is the combination of the linear congruential method and permutation of bit string and has advantages in the aspect of memory (that is, required qubit number) and computational load (that is, circuit depth) compared to other types of PRNG, for example Mersenne Twister \cite{MT}.
It is possible to construct the quantum gate which progresses the PCG sequence as we present later.

We also consider application to concrete problems.
The first one is credit risk measurement, which is mentioned above.
We later present the quantum circuit which calculates sampled values of loss of a credit portfolio using PRNG.
The second one is the integration of a simple multi-variable function, that is, a trigonometric function whose phase depends on two variables.
We consider this for demonstrative purpose and present not only the circuit but also the numerical result calculated by a simulator.

The rest of this paper is organized as follows.
Section \ref{sec:overview} explains the overview of our idea.
Section \ref{sec:implPRNG} presents concrete implementation of the gate which realizes PCG.
In section \ref{sec:appCreRisk} and \ref{sec:demo} we consider application to credit risk measurement and a simple integration, respectively.
Section \ref{sec:ConclAndDisc} contains conclusion and discussion on some issues.
Especially, we discuss the trade-off between qubit number and circuit depth and importance of such a memory-speed trade-off on adjustment of calculation configuration, which will be often necessary when large-scale Monte Carlo by quantum computer is in practical operation in the future.

\section{\label{sec:overview}Overview of the Idea: Quantum Algorithm for Monte Carlo Using Pseudo-random Number}
\subsection{Our Idea}

Applications of the quantum algorithm for Monte Carlo to high-dimensional integration in financial problems can be found in previous works, especially credit risk measurement in \cite{EggerEtAl}.
In the paper, independent random numbers necessary to obtain a value of integrand are represented by different quantum registers, so the number of required qubits $N_{\rm{qubit}}$ is proportional to the number of random numbers $N_{\rm{ran}}$.
If $N_{\rm{ran}}$ is large as in the aforementioned cases, this can lead to shortage of qubits.

Let us see the method in more detail.
The way to represent a random number by quantum register is as follows.
For example, a qubit with state $\sqrt{1-p}\ket{0}+\sqrt{p}\ket{1}$ can be seen as a Bernoulli random number taking 1 with probability $p$.
We can also represent a discretized approximation of a continuous random number like a normal random number on a quantum register \cite{Grover}.
Then, referring to these registers, the value of the integrand is computed into another register and its expectation value is estimated by methods such as \cite{Montanaro, Suzuki}.
Note that this procedure intends to make a superposition of all possible patterns of random numbers\footnote{If a continuous random number is approximated discretely, `all patterns' means those of the discretized value.} and the integrand value and estimate the {\it exact} expectation value, which we hereafter write as $E_{\rm{true}}$.

In order to perform Monte Carlo enjoying quantum speed-up and reducing qubits, we first note that what we calculate in classical Monte Carlo is different from that in the quantum way.
That is, we do {\it not} consider {\it all patterns} of random number values in the classical Monte Carlo.
We {\it sample only a part of patterns} of the random numbers and the integrand and take a simple arithmetic average of the sampled integrand values as an approximation for $E_{\rm{true}}$.
In other words, we calculate $E_{\rm{samp}}$, the expectation value under the sample space which consists of a part of samples and the probability measure under which equal probability is allocated to each sample.
Besides, in most cases, we use a sequence of PRN on behalf of random numbers to calculate the integrand, since strict randomness is difficult to realize on a classical computer.
More concretely, we usually generate a PRN sequence with $N_{\rm{samp}}N_{\rm{ran}}$ elements and divide them into $N_{\rm{samp}}$ subsequences with $N_{\rm{ran}}$ elements, then use each subsequence to calculate a sample value of the integrand\footnote{Mathematically, using such subsequences might raise a statistical concern for large $N_{\rm ran}$, in terms of homogeneity of the distribution of tuples of consecutive PRNs in a high dimensional space \cite{MT}. However, such a way to use PRN is often adopted in practice in banks. We consider that using a tiny part in a large period PRN mitigates the concern \cite{Glasserman}.}.

Our idea is that we estimate not $E_{\rm{true}}$ but $E_{\rm{samp}}$ using a quantum computer in the way similar to classical Monte Carlo.
Before we describe the calculation flow in this method, let us state two assumptions necessary for it.
The first assumption is that on the integrand. 
We assume that it takes $N_{\rm{ran}}$ random numbers as arguments and is sequentially computed in $N_{\rm{ran}}$ steps, each of which requires a random number and the output of the previous step as inputs.
That is, using the intermediate functions $f_n,n=1,...,N_{\rm ran}-1$, the value of the integrand $f_{N_{\rm{ran}}}$ for a given sequence $x_1,...,x_{N_{\rm ran}}$ is calculated as
\begin{eqnarray}
y_1 & = & f_1(x_1), \nonumber \\
y_n & = & f_n(y_{n-1}, x_n) \quad {\rm for} \quad n=2,...,N_{\rm ran}. \label{integrandForm}
\end{eqnarray}
We also assume that $f_{N_{\rm{ran}}}$ is normalized so that $0\le f_{N_{\rm{ran}}}(y,x)\le 1$ for any $x,y$.
Second, we assume that we can choose some PRNG which consumes $n_{\rm{PRN}}$ bits, including the random number itself and working space, and construct two types of quantum gate.
One is $P_{\rm{PRN}}$, which progresses a PRN sequence by one step, that is\footnote{Here and hereafter, a subscript of a ket basically denotes the qubit number of the register.}, 
\begin{equation}
\ket{x_n}_{n_{\rm{PRN}}} \rightarrow \ket{x_{n+1}}_{n_{\rm{PRN}}},
\end{equation}
where $x_n$ is the $n$-th element of the PRN sequence.
The other is $J_{\rm{PRN}}$, which gives $x_{iN_{\rm ran}+1}$ for given $i$, that is,
\begin{equation}
\ket{i}_{n_{\rm{samp}}}\ket{0}_{n_{\rm{PRN}}} \rightarrow \ket{i}_{n_{\rm{samp}}}\ket{x_{iN_{\rm ran}+1}}_{n_{\rm{PRN}}},
\end{equation}
where $n_{\rm{samp}}$ is an integer which satisfies $0<n_{\rm{samp}}<n_{\rm{PRN}}$.
We show a concrete example of PRNGs which satisfies this assumption in section \ref{sec:implPRNG}.

Then, the calculation flow in our method is as follows.
We take $N_{\rm{samp}}$, the number of samples, as $N_{\rm{samp}}=2^{n_{\rm{samp}}}$ for simplicity,

\begin{enumerate}
	\item Prepare a register $R_{\rm{samp}}$ with $n_{\rm{samp}}$ qubits and generate a superposition of $\ket{0}, \ket{1},...,\ket{N_{\rm{samp}}-1}$ with equal amplitudes, that is, $\frac{1}{\sqrt{N_{\rm{samp}}}}\sum_{i=0}^{N_{\rm{samp}}-1}{\ket{i}_{n_{\rm{PRN}}}}$. This can be done by operating a Hadamard gate to each of the $n_{\rm{samp}}$ qubits.
	\item Operate $J_{\rm{PRN}}$, then the $(iN_{\rm{ran}}+1)$-th element of the sequence is set to the register $R_{\rm{PRN}}$, where $i$ is determined by the state of $R_{\rm{samp}}$. These are the starting points of subsequences.
	\item Perform a calculation step of the integral referring to $R_{\rm{PRN}}$ and reflect the result into a register $R_{\rm int}$. Here, we assume that the integrand is calculated step-by-step using each random number.
	\item Operate $P_{\rm{PRN}}$ to the register $R_{\rm{PRN}}$, then the PRN sequence progresses by one step.
	\item Perform a calculation step of the integral referring to $R_{\rm{PRN}}$ and reflect the result into a register $R_{\rm int}$.
	\item Iterate 4 and 5 until the calculation of the integrand ends. This corresponds to sequential generation of PRN and calculation using it. Finally, we obtain an equiprobable superposition of states, in each of which $R_{\rm{int}}$ holds a sampled integrand value.
	\item Prepare an ancilla qubit, which we call $R_{\rm ph}$, and encode the integrand value into the amplitude of $\ket{1}$ in $R_{\rm ph}$ using controlled rotations.
	\item Estimate the amplitude of the state where $R_{\rm ph}$ is $\ket{1}$ by the amplitude estimation methods like \cite{Bassard,Suzuki}. This is an estimate for the arithmetic average of sampled integrand values, that is, $E_{\rm{samp}}$.
\end{enumerate}

\begin{figure*}[t]
	\begin{minipage}{1\hsize}
		\begin{center}
			\includegraphics[width=1\textwidth]{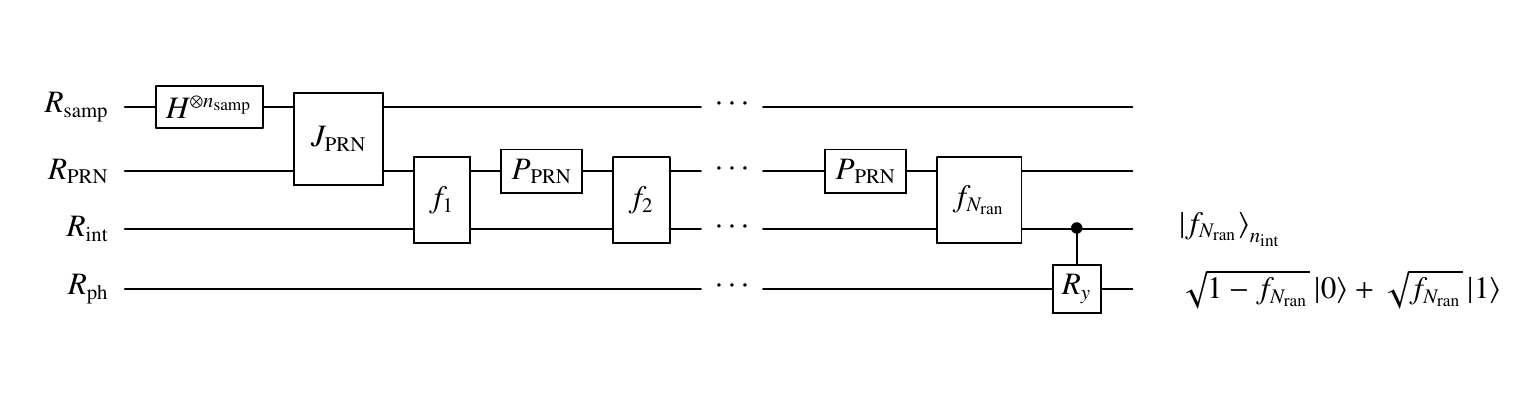}
		\end{center}
		\subcaption{}
		\label{OverviewCircuitOurs}
	\end{minipage}\\	
	
	\begin{minipage}{1\hsize}
		\begin{center}
			\includegraphics[width=0.7\textwidth]{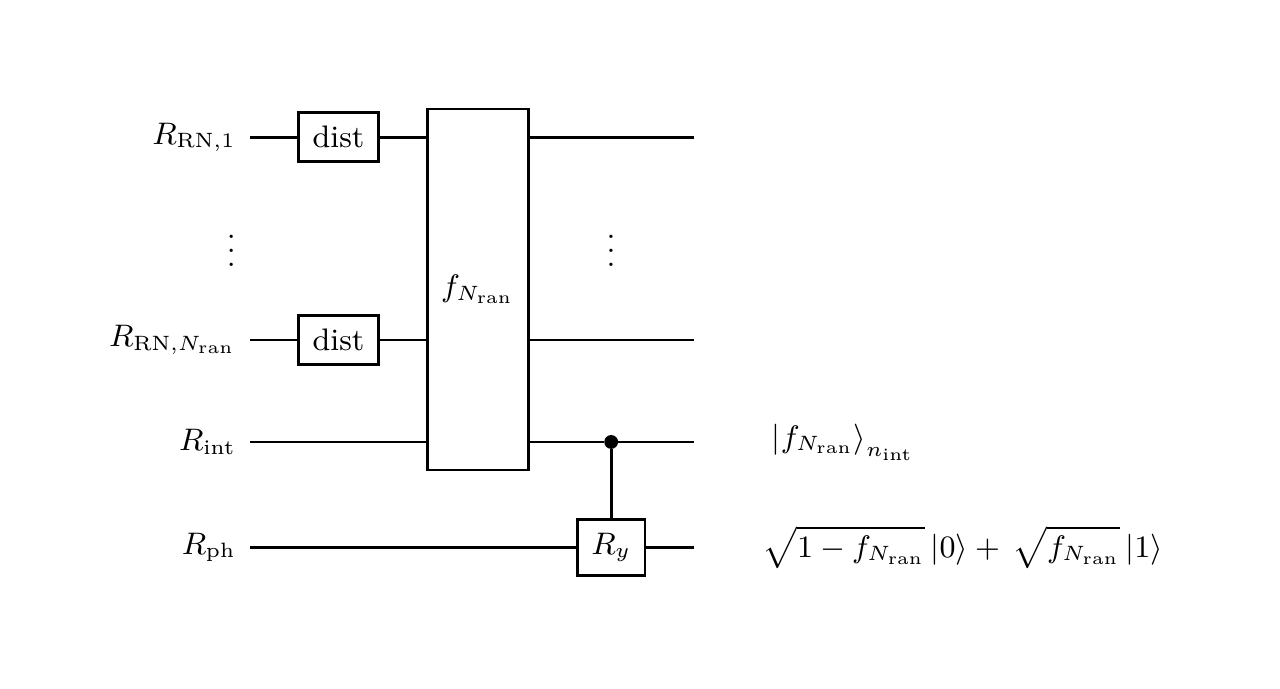}
		\end{center}
		\subcaption{}
		\label{OverviewCircuitPrev}
	\end{minipage}	
	\caption{Circuits for the quantum algorithm for Monte Carlo. Figures (a) and (b) correspond to the method we propose and that in previous papers, respectively. These are overviews and ancillas are not shown. Actually, the circuit for amplitude estimation follows the above circuits, but we omit it. See \cite{Bassard,Suzuki} for the detail.}
	\label{OverviewCircuit}
\end{figure*}

The flow of state transformation on $R_{\rm{samp}},R_{\rm{PRN}},R_{\rm{int}}$ and $ R_{\rm ph}$ is as follows:

\begin{widetext}
\begin{eqnarray}
& & \ket{0}_{n_{\rm{samp}}}\ket{0}_{n_{\rm{PRN}}}\ket{0}_{n_{\rm{int}}}\ket{0}  \nonumber \\
& \xrightarrow{1} & \frac{1}{\sqrt{N_{\rm{samp}}}}\sum_{i=0}^{N_{\rm{samp}}-1} {\ket{i}_{n_{\rm{samp}}}}\ket{0}_{n_{\rm{PRN}}}\ket{0}_{n_{\rm{int}}}\ket{0} \nonumber \\
& \xrightarrow{2} & \frac{1}{\sqrt{N_{\rm{samp}}}}\sum_{i=0}^{N_{\rm{samp}}-1} {\ket{i}_{n_{\rm{samp}}} \ket{x^{(i)}_1}_{n_{\rm{PRN}}}}\ket{0}_{n_{\rm{int}}}\ket{0} \nonumber \\
&\xrightarrow{3} & \frac{1}{\sqrt{N_{\rm{samp}}}}\sum_{i=0}^{N_{\rm{samp}}-1}  {\ket{i}_{n_{\rm{samp}}}\ket{x^{(i)}_1}_{n_{\rm{PRN}}}}\ket{f^{(i)}_1}_{n_{\rm{int}}}\ket{0} \nonumber \\
&\xrightarrow{4} & \frac{1}{\sqrt{N_{\rm{samp}}}}\sum_{i=0}^{N_{\rm{samp}}-1}  {\ket{i}_{n_{\rm{samp}}} \ket{x^{(i)}_2}_{n_{\rm{PRN}}}}\ket{f^{(i)}_1}_{n_{\rm{int}}}\ket{0} \nonumber \\
&\xrightarrow{5} & \frac{1}{\sqrt{N_{\rm{samp}}}}\sum_{i=0}^{N_{\rm{samp}}-1} {\ket{i}_{n_{\rm{samp}}}\ket{x^{(i)}_2}_{n_{\rm{PRN}}}}\ket{f^{(i)}_2}_{n_{\rm{int}}}\ket{0} \nonumber \\
&\xrightarrow{6} & ... \qquad\quad\quad\quad  \nonumber \\
&\xrightarrow{6} & \frac{1}{\sqrt{N_{\rm{\rm{samp}}}}}\sum_{i=0}^{N_{\rm{\rm{samp}}}-1}  {\ket{i}_{n_{\rm{samp}}}\ket{x^{(i)}_{N_{\rm{ran}}}}_{n_{\rm{PRN}}}} \ket{f^{(i)}_{N_{\rm{ran}}}}_{n_{\rm{int}}}\ket{0} \nonumber \\
&\xrightarrow{7} & \frac{1}{\sqrt{N_{\rm{\rm{samp}}}}}\sum_{i=0}^{N_{\rm{\rm{samp}}}-1}  {\ket{i}_{n_{\rm{samp}}}\ket{x^{(i)}_{N_{\rm{ran}}}}_{n_{\rm{PRN}}}} \ket{f^{(i)}_{N_{\rm{ran}}}}_{n_{\rm{int}}}\left(\sqrt{1-f^{(i)}_{N_{\rm{ran}}}}\ket{0}+\sqrt{f^{(i)}_{N_{\rm{ran}}}}\ket{1}\right)  \label{overallflow}
\end{eqnarray}
\end{widetext}
Here, $x^{(i)}_n=x_{iN_{\rm{ran}}+n}$ and this is the $n$-th element of the $i$-th subsequence.
$n_{\rm{int}}$ is the qubit number of $R_{\rm{int}}$.
$f^{(i)}_1,..,f^{(i)}_{N_{\rm{ran}}}$ are the values of $f_1,...,f_{N_{\rm{ran}}}$ for $x^{(i)}_1,...,x^{(i)}_{N_{\rm ran}}$.
%$f_2(f_1(x_1), x_2)$ is simply written as $f_2(x_1, x_2)$, $f_3(f_2(x_1, x_2), x_3)$ as $f_3(x_1, x_2, x_3)$, and so on.

We present an outline of the quantum circuit for the above method in Figure \ref{OverviewCircuit}.
We also present that for the method in the previous papers for comparison.
In our method, as shown in Figure \ref{OverviewCircuitOurs}, after the operation which create a superposition of $\ket{x^{(0)}_1},\ket{x^{(1)}_1},...,\ket{x^{(N_{\rm samp}-1)}_1}$ on $R_{\rm{PRN}}$ and the gate $f_1$, the first step of calculation of the integrand, we sequentially operate $P_{\rm{PRN}}$ and $f_n$, the $n$-th calculation step.
The register which represents (pseudo) random numbers is only $R_{\rm{PRN}}$ and PRNs $x^{(i)}_n$ are sequentially generated on it.
The intermediate value of the integrand $f^{(i)}_{n}$ is calculated into $R_{\rm{int}}$ using $x^{(i)}_n$ and $f^{(i)}_{n-1}$ as inputs and finally $f^{(i)}_{N_{\rm{ran}}}$ is reached.
On the other hand, in the method in previous works, as shown in Figure \ref{OverviewCircuitPrev}, quantum registers $R_{{\rm RN},1},...,R_{{\rm RN},N_{\rm ran}}$ are prepared to represent all random numbers simultaneously and a superposition of numbers following the desired probability distribution (for example, normal) is generated on each register by the gate 'dist' in Figure \ref{OverviewCircuitPrev}.
Then, the integrand value is calculated using all of $R_{{\rm RN},1},...,R_{{\rm RN},N_{\rm ran}}$ at the same time.

Here we make some comments.
The first one is about the probability distribution of random numbers.
In the previous method, a random number under the desired distribution is generated on a register using the gate 'dist'.
On the other hand, in the method of this paper, sequentially generated PRNs basically obey uniform distribution, since most PRNGs are for that distribution.
Therefore, we have to convert uniform random numbers to random numbers obeying a desired distribution.
Such a conversion is actually a common step in the classical Monte Carlo and there are many well-known methods, for example, the Box-Muller method for standard normal distribution.
We assume such a conversion is implementable as a quantum gate and contained in $f_n$.
Actually, implementation of trigonometric functions and logarithm, which are necessary to the Box-Muller method, has been investigated in previous papers \cite{Haner,Cao,Bhaskar}.

\begin{figure*}[tb]
	%\begin{center}
	\includegraphics[width=0.8\textwidth]{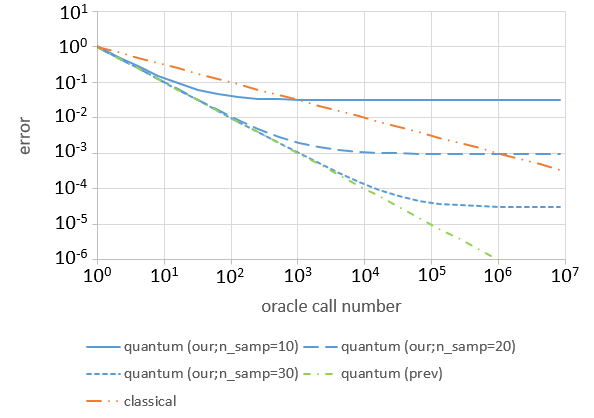}
	%\end{center}
	\caption{Errors in various methods for Monte Carlo. The blue solid, dashed and dotted lines are $\Delta_{\rm our}$ in (\ref{eq:DeltaOur}), the errors in the quantum method which we propose in this paper for various values of $n_{\rm samp}$. The solid, dashed and dotted lines correspond to $n_{\rm samp}=10,20$ and 30, respectively. The green chain line is $\Delta_{\rm prev}$ in (\ref{eq:DeltaPrev}), the error in the quantum methods in the previous papers. The red two-dot chain line is $\Delta_{\rm class}$ in (\ref{eq:DeltaClass}), the error in the classical method. The horizontal axis is the oracle call number $N_{\rm orac}$. We here set $c\sigma_{f_{N_{\rm ran}}}=1,2\pi d \sqrt{E_{\rm samp}(1-E_{\rm samp})}=1,2\pi d \sqrt{E_{\rm true}(1-E_{\rm true})}=1$.}
	\label{ErrorGraph}
\end{figure*}

The second comment is about how the distribution of the integrand value is taken into account in the method of this paper.
In the previous method, the desired distribution of the integrand value is realized through the distribution of random numbers on registers.
On the other hand, the distribution of the integrand value is generated through the PRNG in the method of this paper.
Although the final state is a superposition of various integrand values with equal probability, the appearance pattern of the value reflects the distribution.
For example, if the integrand value obeys distribution with a peak at some value $F$, the integrand values close to $F$ frequently appear on $R_{\rm{int}}$ in the set of states that compose the final superposition.

Finally, we comment on the integrand form such that it can be computed sequentially.
As mentioned above, it is just an assumption, that is, not all functions can be written like this.
However, in many cases, the integrand takes this form; for examples of use cases of Monte Carlo, see textbooks such as \cite{Glasserman}.
We here give two frequent cases which satisfy (\ref{integrandForm}) and include some important problems.
First, the integrand takes the form of (\ref{integrandForm}) if, after fixing some random numbers, we can write contributions from remaining random numbers in a separable sum or product, that is, $f_{N_{\rm{ran}}}(x_1,...,x_{N_{\rm ran}}) = \sum_{j=d+1}^{N_{\rm{ran}}}g_j(x_j;x_1,...,x_d)$ or $f_{N_{\rm{ran}}}(x_1,...,x_{N_{\rm ran}}) = \prod_{j=d+1}^{N_{\rm{ran}}}g_j(x_j;x_1,...,x_d)$, where $d$ is a small natural number compared with $N_{\rm{ran}}$ and $g_j,j=d+1,...,N_{\rm{ran}}$ are some functions.
The credit portfolio risk measurement, which we will consider later, corresponds to this case.
Second, when we simulate Markov processes, they can be calculated in the sequential way like above.
Pricing of financial derivative, where the underlying assets are Markov in many cases, is a typical example of this.
Thinking of these examples, we are well motivated to consider the case where (\ref{integrandForm}) is satisfied.

\subsection{Relationship between Computational Load and Error}

Now, we roughly estimate the relationship between computational load and additive error in three methods: the quantum method we propose, the quantum method proposed in previous papers and the classical method.
In this paper, we measure computational load by $N_{\rm orac}$, the number of times that we call the {\it oracle} in each method. Here, the {\it oracle} means the procedure to calculate the integrand. More concretely, it is the circuit (that in Figure \ref{OverviewCircuit}, in the current case) and the subroutine to calculate the integrand for the quantum and classical method, respectively. In fact, in quantum methods, we have to repeatedly call the oracle circuit for amplitude estimation with the desired error level and this occupies the dominant part of the computation. On the other hand, the classical method requires the sufficient number of sampling to reduce the error and the computational time is almost proportional to the sample number.

First, let us consider our method.
There are two sources of error.
One is the difference between $E_{\rm samp}$ and $E_{\rm true}$, which we write as $\Delta_{\rm TrSm}$.
The other is the estimation error of $E_{\rm samp}$, that is, the error of amplitude estimation, which we write as $\Delta_{\rm Est}$.
$\Delta_{\rm TrSm}$ is equal to the statistical error in the classical Monte Carlo.
For some fixed confidence level, it is at most $c\sigma_{f_{N_{\rm ran}}}N_{\rm samp}^{-1/2}=c\sigma_{f_{N_{\rm ran}}}2^{-n_{\rm{samp}}/2}$, where $\sigma_{f_{N_{\rm ran}}}$ is the standard deviation of $f_{N_{\rm ran}}$ and $c$ is a constant set according to the confidence level.
Note that it depends on not $N_{\rm orac}$ but $n_{\rm{samp}}$.
On the other hand, $\Delta_{\rm Est}$ is estimated as follows.
The quantum algorithm in \cite{Bassard} with $N_{\rm orac}$ oracle calls gives estimation for the amplitude (that is, $E_{\rm samp}$) which differs from the true value by at most $2\pi d \sqrt{E_{\rm samp}(1-E_{\rm samp})}N_{\rm orac}^{-1}$ with probability at least $1-\delta$.
Here, we take only the leading term with respect to $N_{\rm orac}^{-1}$ and $d$ is some $O(1)$ constant depending only on $\log\delta$.
In total, the error in our method is at most
\begin{eqnarray}
\Delta_{\rm our} & \sim & \Delta_{\rm TrSm} + \Delta_{\rm Est} \nonumber \\
& \simeq & c\sigma_{f_{N_{\rm ran}}}2^{-n_{\rm{samp}}/2}+2\pi d \sqrt{E_{\rm samp}(1-E_{\rm samp})}N_{\rm orac}^{-1}. \label{eq:DeltaOur}
\end{eqnarray}
If we desire the error level $\epsilon$, the following setting is sufficient.
First, we set
\begin{equation}
N_{\rm samp}\sim \left(\frac{c\sigma_{f_{N_{\rm ran}}}}{\epsilon}\right)^2,
\end{equation}
or, equivalently,
\begin{equation}
n_{\rm samp}\sim \lceil 2\log_2{(c\sigma_{f_{N_{\rm ran}}}/{\epsilon})}\rceil,
\end{equation}
so that $\Delta_{\rm TrSm}\sim\epsilon$. Then, we set
\begin{equation}
N_{\rm orac} \sim \frac{2\pi d \sqrt{E_{\rm samp}(1-E_{\rm samp})}}{\epsilon},
\end{equation}
which leads to $\Delta_{\rm Est}\sim\epsilon$. 
Note that $N_{\rm orac}$ does not depend on $N_{\rm samp}$.

This is actually quadratic speed up compared with the classical Monte Carlo.
In the classical method, the error is
\begin{equation}
\Delta_{\rm class} \sim c\sigma_{f_{N_{\rm ran}}}N_{\rm orac}^{-1/2}, \label{eq:DeltaClass}
\end{equation}
as $\Delta_{\rm TrSm}$ in (\ref{eq:DeltaOur}). Note that $N_{\rm orac}=N_{\rm samp}$ for the classical method. Then, the required $N_{\rm orac}$ in the method is
\begin{equation}
N_{\rm orac}\sim\left(\frac{c\sigma_{f_{N_{\rm ran}}}}{\epsilon}\right)^2
\end{equation}
for the desired error $\epsilon$.

We also mention the error in the previous method of the quantum-based Monte Carlo.
This method estimates $E_{\rm true}$ itself and the error is at most
\begin{equation}
\Delta_{\rm prev} \sim 2\pi d \sqrt{E_{\rm true}(1-E_{\rm true})}N_{\rm orac}^{-1}, \label{eq:DeltaPrev}
\end{equation}
for the oracle call number $N_{\rm orac}$.
Note that this estimated error is nearly equal to $\Delta_{\rm Est}$ in (\ref{eq:DeltaOur}).
This is because the error of amplitude estimation is determined by the amplitude itself and $N_{\rm orac}$ only \cite{Bassard} and the estimated amplitude is almost equal in both of the previous and our methods as long as $\Delta_{\rm TrSm}$ is small.

Figure \ref{ErrorGraph} represents the relationship among these errors in some specific case.
We plot $\Delta_{\rm our}, \Delta_{\rm prev}$ and $\Delta_{\rm class}$ versus $N_{\rm orac}$.
We here set the prefactors $c\sigma_{f_{N_{\rm ran}}},2\pi d \sqrt{E_{\rm samp}(1-E_{\rm samp})},2\pi d \sqrt{E_{\rm true}(1-E_{\rm true})}$ to 1. Besides, we set $n_{\rm samp}=10,20,30$, which correspond to $N_{\rm samp}=2^{10}(\approx 10^3),2^{20}(\approx 10^6),2^{30}(\approx 10^9)$, respectively.
$N_{\rm samp}=10^6$ is a typical value in the case of the credit risk measurement.
When $\Delta_{\rm TrSm} \ll \Delta_{\rm Est},$
$\Delta_{\rm our}$ decreases faster than $\Delta_{\rm class}$ and similar to $\Delta_{\rm prev}$ as $N_{\rm orac}$ increases.
After $\Delta_{\rm Est}$ becomes smaller than $\Delta_{\rm TrSm}$, $\Delta_{\rm our}$ asymptotically converges to $\Delta_{\rm TrSm}$.
However, for sufficiently large $n_{\rm samp}$, $\Delta_{\rm our}$ reaches the same order of magnitude as $\Delta_{\rm class}$ for smaller $N_{\rm orac}$.
For example, when $n_{\rm samp}\ge 20$, $\Delta_{\rm our}$ reaches the same order of magnitude as $\Delta_{\rm class}$ for $N_{\rm orac}=10^6$ only for $N_{\rm orac}=10^3$, smaller by three orders of magnitude.
In such a region, our method has an advantage compared to the classical way.

We also note that increasing $n_{\rm{samp}}$ by a few leads to increase of $N_{\rm samp}$ and decrease of $\Delta_{\rm TrSm}$ by orders of magnitude.
$N_{\rm ran}N_{\rm samp}$ cannot exceed $P$, the period of PRN, but we expect that it is unnecessary to concern such a upper bound, as long as we use a widely-used PRNG, which has a period, say $2^{64}$.
At least in the case of the credit risk measurement, this is much longer than $N_{\rm ran}N_{\rm samp}$ in practice, since each of these is at most $10^6$ and the product is at most $10^{12} \sim 2^{40}$.

\section{\label{sec:implPRNG}Implementation of Pseudo-random Number Generator on Quantum Circuit}

\begin{figure*}[t]
	\begin{center}
		\begin{tabular}{c}
			
			% 1
			\begin{minipage}{0.5\hsize}
				\begin{center}
					\includegraphics[width=1\textwidth]{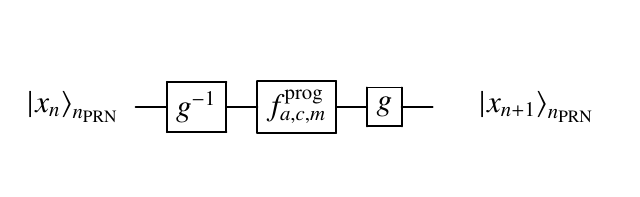}
				\end{center}
				\subcaption{}
			\end{minipage}
			
			% 2
			\begin{minipage}{0.5\hsize}
				\begin{center}
\includegraphics[width=1\textwidth]{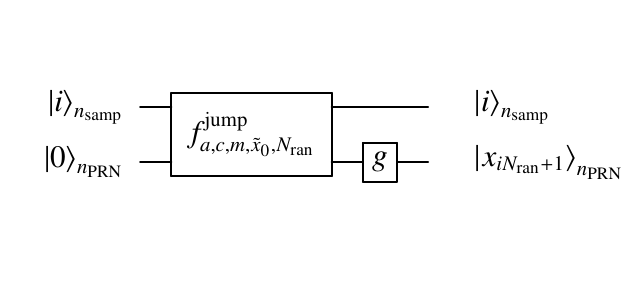}
				\end{center}
				\subcaption{}
			\end{minipage}
			
		\end{tabular}
		\caption{Quantum gates for PCG. Figures (a) and (b) correspond to $P_{\rm PRN}$ and $J_{\rm PRN}$, respectively.}
		\label{PCGCircuit}
	\end{center}
\end{figure*}

\begin{figure*}[t]
	\begin{center}
		\begin{tabular}{c}
			
			\begin{minipage}{0.5\hsize}
				\begin{center}
\includegraphics[width=1\textwidth]{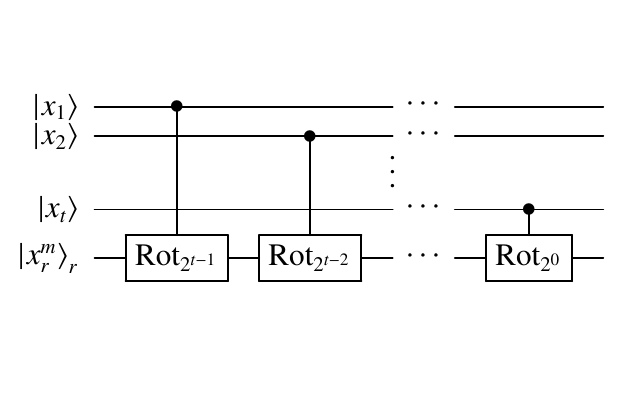}
				\end{center}
				\subcaption{}
			\end{minipage}
			
			\begin{minipage}{0.5\hsize}
				\begin{center}
				\includegraphics[width=1\textwidth]{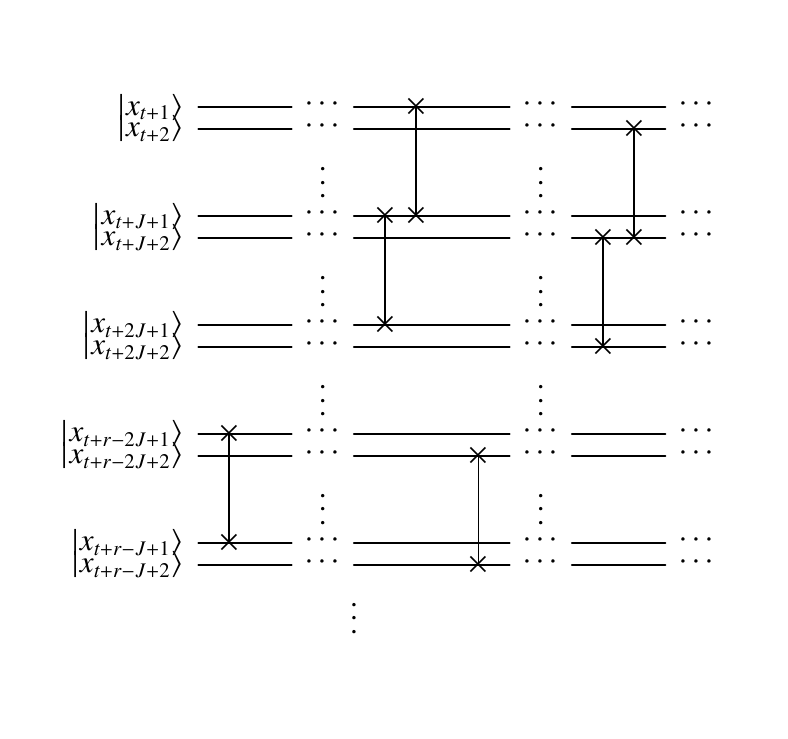}	
				\end{center}
				\subcaption{}
			\end{minipage}
			
		\end{tabular}
		\caption{Quantum gate which performs random rotation. Figure (a) is the overview and (b) is the detail of ${\rm Rot}_J,J=2^j$.}
		\label{RandomRot}
	\end{center}
\end{figure*}

\begin{figure*}[t]
	\begin{center}
		\includegraphics[width=0.4\textwidth]{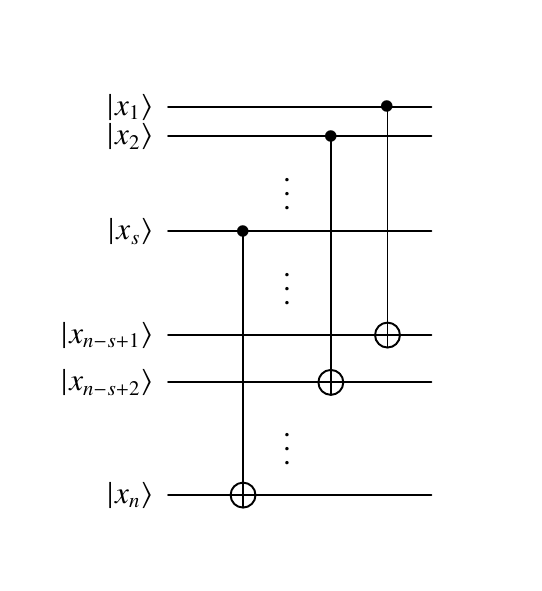}
		\caption{Quantum gate which performs xorshift.}
		\label{Xorshift}
	\end{center}
\end{figure*}

\subsection{PCG}
We next consider how to implement a PRNG on a quantum circuit, that is, the gates $P_{\rm PRN}$ and $J_{\rm PRN}$.
Remembering the motivation of this work, reduction of qubits, PRNGs which require small working space are desirable.
Besides, in order to decrease circuit depth as much as possible, we desire a simpler and shorter calculation step to progress PRN sequence.
Of course, the longer period and better statistical property is preferred.
We propose PCG \cite{PCG} as a PRNG which satisfies these properties.

PCG is combination of linear congruential generator (LCG), a popular and elementary PRNG, and permutation of bit string.
The $n$-th element of a PCG sequence $x_n$ is recursively defined as follows:
\begin{eqnarray}
\begin{cases}
\tilde{x}_{n+1} = f^{\rm prog}_{a,c,m}(\tilde{x}_{n}) \coloneqq (a\tilde{x}_{n}+c) \mod m & \\
x_n=g(\tilde{x}_{n}), &
\end{cases}
\end{eqnarray}
where $a,c$ and $m$ are integer parameters satisfying $a>0, c \geq 0, m>0$ and the seed $\tilde{x}_0$ is also given as an integer satisfying $0\leq \tilde{x}_0 < m$.
$\tilde{x}_{n}$ is the background sequence and defined by the LCG recurrence formula as above.
$g$ is the permutation of a bit string, which is explained in detail later.
Therefore, the calculation steps to progress a PCG sequence is the sequence of modular multiplication, modular addition and permutation.
Besides, for LCG we can easily jump ahead by $k$ steps using the following formula:
\begin{equation}
\tilde{x}_{n+k} = \left(a^k\tilde{x}_{n}+\frac{c(a^k-1)}{a-1}\right) \mod m.
\end{equation}
Especially, we can obtain $\tilde{x}_{iN_{\rm ran}+1}$ from a seed $x_0$ as 
\begin{equation}
\tilde{x}_{iN_{\rm ran}+1} = f^{\rm jump}_{a,c,m,\tilde{x}_0,N_{\rm ran}}(i) \coloneqq \left(a^{iN_{\rm ran}+1}\tilde{x}_0+\frac{c(a^{iN_{\rm ran}+1}-1)}{a-1}\right) \mod m.
\end{equation}

Given the above formulas, we can construct quantum gates $P_{\rm PRN}$ and $J_{\rm PRN}$ for PCG.
The rough images of the circuit diagrams are shown in Figure \ref{PCGCircuit}.
To construct $P_{\rm PRN}$, we first get back PCG to LCG with the inverse of $g$, then progress LCG with the $f^{\rm prog}_{a,c,m}$ gate and finally perform the permutation $g$.
The $f^{\rm prog}_{a,c,m}$ gate maps $\ket{x}$ to $\ket{f^{\rm prog}_{a,c,m}(x)}$ and is constructed as modular multiplication $\ket{x} \rightarrow \ket{ax \mod m}$ followed by modular addition $\ket{x} \rightarrow \ket{(x + c) \mod m}$.
To construct $J_{\rm PRN}$, we first operate the $f^{\rm jump}_{a,c,m,\tilde{x}_0,N_{\rm ran}}$ gate, which refers to the first register as an input and transforms the second register from $\ket{0}$ to $\ket{\tilde{x}_{iN_{\rm ran}+1}}$ if the first register is $\ket{i}$, then $g$ to the second register.
The  $f^{\rm jump}_{a,c,m,\tilde{x}_0,N_{\rm ran}}$ gate is constructed as a combination of modular addition, subtraction, multiplication, division and exponentiation $\ket{k}\ket{x} \rightarrow \ket{k}\ket{a^kx \mod m}$.
Implementation of (modular) adder, multiplier and exponentiator has been investigated in many papers, for example, \cite{Vedral,Beckman,Draper,Cuccaro,Takahashi,VanMeter,Draper2,Takahashi2,Portugal,AlvarezSanchez,Takahashi3,Thapliyal}.
Modular subtraction is the inverse of addition.
Division by $a-1$ modulo $m$ is implemented as multiplication by an integer $b$ such that $(a-1)b\equiv 1 \mod m$, which can be found by the extended Euclidean algorithm\footnote{Such $b$ can be found if and only if $a-1$ and $m$ are coprime. This condition is satisfied for many of widely used combination of $a$ and $m$.}\cite{Knuth}.

There is a comment on implementation of $f^{\rm prog}_{a,c,m}$.
It should be implemented not in the form that it output the result in the register other than the input register, that is, $\ket{x}\ket{0} \rightarrow \ket{x}\ket{f^{\rm prog}_{a,c,m}(x)}$, but in the form that it updates the input register itself into the resulting state, that is, $\ket{x} \rightarrow \ket{f^{\rm prog}_{a,c,m}(x)}$.
This is because this gate is repeatedly used in the method of this paper, so the qubit number required for the entire calculation explodes if it is necessary to add a register in each calculation step.
Most of the previous implementation of modular addition are the self-updating type, and so can be used with no change.
On the other hand, some implementation of modular multiplication output the result into ancilla, $\ket{x}\ket{0} \rightarrow \ket{x}\ket{ax \mod m}$, but the trick described in \cite{Markov} solves the problem as follows.
First, using an integer $a^{\prime}$ such that $aa^{\prime}\equiv 1 \mod m$, we construct a gate which performs $\ket{x}\ket{0} \rightarrow \ket{x}\ket{a^{\prime}x \mod m}$ and its inverse.
Then, we can implement the following sequence:
\begin{eqnarray}
\ket{x}\ket{0} & \rightarrow & \ket{x}\ket{ax \mod m} \nonumber \\
& \rightarrow & \ket{ax \mod m}\ket{x} \nonumber \\
& \rightarrow &  \ket{ax \mod m}\ket{0}.
\end{eqnarray}
Here, the first, second and third steps are modular multiplication by $a$, swap and the inverse of modular multiplication by $a^\prime$, respectively.

\subsection{Permutation}

It is well-known that LCG suffers from some statistical flaws.
\cite{PCG} points out that performing permutation on LCG enhances its statistical properties.
Here, permutation is transformation of binary representation of a PRN to another bit string.
We here take some of the permutations described in \cite{PCG} as examples and show how to implement them in a quantum circuit.

The first one is random rotation.
We first divide a $n$-bit binary $x\in\mathbb{Z}_{2^n}$ into three parts: the top $t$ bits $x_t^h$, the middle $r$ bits $x_r^m$ and the bottom $n-t-r$ bits $x_{n-t-r}^b$, where $r$ is a power of 2 and $t=\log_2 r$.
Then random rotation is a map from $\mathbb{Z}_{2^n}$ to $\mathbb{Z}_{2^r}$ defined as
\begin{equation}
x \mapsto \sigma_{\rm rot}(x_t^h, x_r^m).
\end{equation}
Here,
\begin{equation}
\sigma_{\rm rot}(k, y)\coloneqq
\begin{cases}
y & ;k=0 \\
y_{r-k+1}...y_{r}y_1...y_{r-k} & ;1\leq k \leq r-1
\end{cases}
\end{equation}
for an integer $k$ satisfying $0\leq k \leq r-1$, $y=y_1y_2...y_r\in \mathbb{Z}_{2^r}$ and $ab...$ represents a bit string whose first digit is $a\in\{0,1\}$, second digit is $b\in\{0,1\}$ and so on.
In short, random rotation is clockwise rotation of middle digits of a binary where the rotation width is determined by the value of top digits.
Only the middle digits $x_r^m$ are used to calculate the integrand as a $r$ bit random number.
Especially, the bottom digits $x_{n-t-r}^b$ are discarded since their statistical properties are not good.

Random rotation is easily implemented in a quantum circuit using controlled swap gate (Fredkin gate).
The circuit diagram is shown in Figure \ref{RandomRot}.
The middle bits $\ket{x^m_r}_r$ is rotated by $2^{t-i}$ bits by ${\rm Rot}_{2^{t-i}}$ under the control of the top $i$-th bit, for $1\leq i \leq t$.
This leads to $x_t^h$-bit rotation of $x_r^m$.
We can construct the gate ${\rm Rot}_{2^j},j=0,1,...,t-1$ connecting qubits with swap gates (actually Fredkin gates since these gates are controlled) as follows.
Setting $J=2^j$,

\begin{itemize}
	\item Connect $\ket{x_{t+(r/J-2)\cdot J+1}}$ and $\ket{x_{t+(r/J-1)\cdot J+1}}$, $\ket{x_{t+(r/J-3)\cdot J+1}}$ and $\ket{x_{t+(r/J-2)\cdot J+1}}$,...,$\ket{x_{t+1}}$ and $\ket{x_{t+J+1}}$
	\item Connect $\ket{x_{t+(r/J-2)\cdot J+2}}$ and $\ket{x_{t+(r/J-1)\cdot J+2}}$,$\ket{x_{t+(r/J-3)\cdot J+2}}$ and $\ket{x_{t+(r/J-2)\cdot J+2}}$,...,$\ket{x_{t+2}}$ and $\ket{x_{t+J+2}}$
	\item ...
	\item Connect $\ket{x_{t+(r/J-2)\cdot J+J}}$ and $\ket{x_{t+(r/J-1)\cdot J+J}}$,$\ket{x_{t+(r/J-3)\cdot J+J}}$ and $\ket{x_{t+(r/J-2)\cdot J+J}}$,...,$\ket{x_{t+J}}$ and $\ket{x_{t+J+J}}$ 
\end{itemize}
That is, there are $J$ groups containing $n/J$ qubits connected by $n/J - 1$ swap gates.
Note that $r/J$ is an integer.

The second type of permutation is xorshift.
This is a map from $\mathbb{Z}_{2^n}$ to $\mathbb{Z}_{2^n}$ defined as follows:
\begin{eqnarray}
x = x_1...x_n \rightarrow x_1...x_{n-s}y_1...y_s, \nonumber \\
y_i \coloneqq x_i \oplus x_{n-s+i}, i=1,...,s.
\end{eqnarray}
Here, $s$ is an integer satisfying $1\leq s\leq n-1$ and typically comparable with $n$, for example, $n/2$ as proposed in \cite{PCG}.
Note that we do not need to take xorshift over the whole qubits in the PRN register.
That is, we can take XOR between top qubits and middle qubits and discard bottom ones, as random rotation.

We can construct a gate which performs this permutation using CNOT gates.
That is, put NOT on $\ket{x_{n-s+i}}$ under control by $\ket{x_i}$, for $i=1,...,s$, as shown in Figure \ref{Xorshift}.
Note the order to set CNOT gates, that is, from bottom to top.
This is necessary in the case where $s > n/2$ so that some middle qubits are used as both a target and a control.
Such a qubit must work as a control before it becomes a target.

\subsection{Qubit Number and Circuit Depth}

Here, we roughly estimate qubit number and depth of PCG circuits.
We focus on $P_{\rm PRN}$, which is repeatedly used to progress PRN sequences.
We consider PCG with $r$-bit output and $n$-bit background LCG.
$n$ should be so large that the period, $2^n$ at most, is long enough and $r$ is typically comparable with $n$.
For example, in many of the settings considered in \cite{PCG}, $n=64$ and $r=32$.

The LCG part consists of modular addition and multiplication and dominant contribution comes from the latter.
For $n$-bit operands, many of the proposed modular adder require $O(n)$ qubits including ancilla and $O(n)$ depth.
On the other hand, modular multipliers basically require $O(n)$ qubits and $O(n^2)$ depth, so this is dominant in the LCG part\footnote{There are implementations which have depth proportional to smaller powers of $n$ than 2 but require ancilla proportional to larger powers of $n$ than 1 \cite{Portugal}.}.

The permutation part does not require any ancillas; at least two examples are mentioned above.
Circuit depth is found as follows.
For random rotation on $r$ bits with $t=\log_2r$ control bits, which we considered above, we first note that depth of swap gates in ${\rm Rot}_{2^j}$ is $r/2^j-1$, since ${\rm Rot}_{2^j}$ consists of $2^j$ groups of $r/2^j$ qubits and $r/2^j-1$ swap gates, as explained above.
Summing this up for $j=0,1,...,t-1$, it is found that the depth of Fredkin gates in the random rotation is $2r-\log_2{r}-2$, that is, $O(r)$.
For xorshift with shift width $s$, it is obvious that the depth of CNOT gates is $s$ and if $s$ is comparable with $r$, say $s=r/2$ as considered in \cite{PCG}, so is the depth.

In summary, in terms of both ancilla qubit number and circuit depth, dominant contribution comes from a multiplication and is $O(n)$ and $O(n^2)$ respectively.
Therefore, if each calculation step for integrand $f_i$ contains computations heavier than several multiplications, PRN generation makes subdominant contributions to qubit number and circuit depth.

\section{\label{sec:appCreRisk}Application to Credit Risk Measurement}

\begin{figure*}[t]
	\begin{minipage}{1\hsize}
		\begin{center}
			\includegraphics[width=1\textwidth]{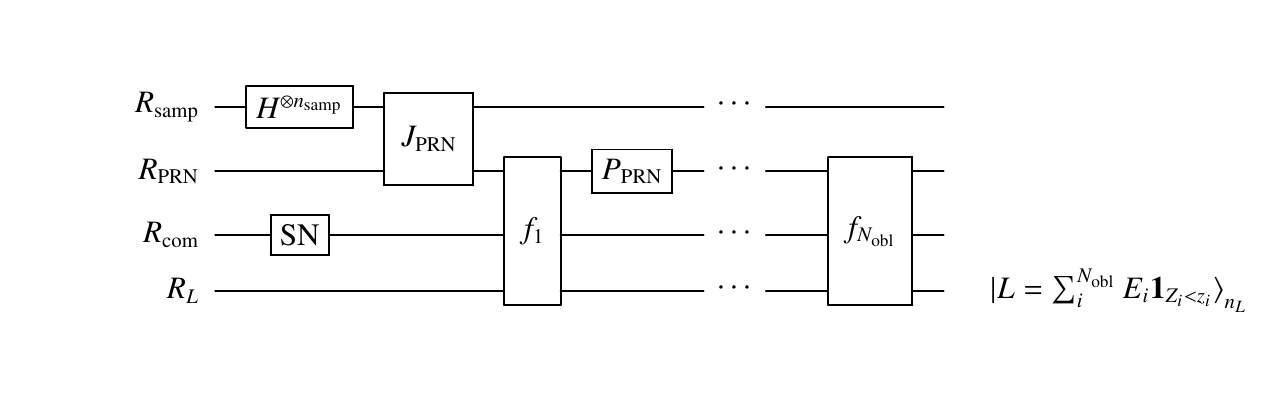}
		\end{center}
		\subcaption{}
	\end{minipage}\\	
	
	\begin{minipage}{1\hsize}
		\begin{center}
			\includegraphics[width=1\textwidth]{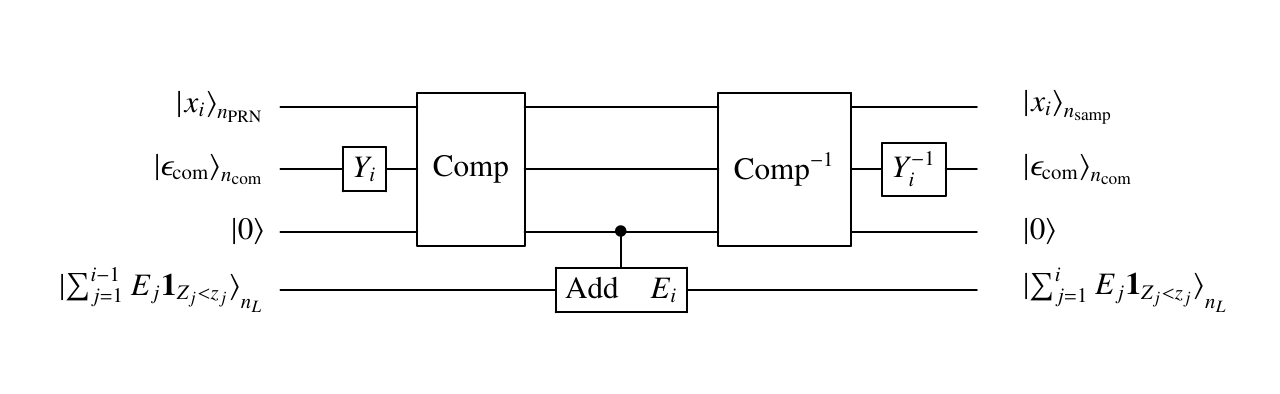}
		\end{center}
		\subcaption{}
	\end{minipage}	
	\caption{The Quantum circuit to calculate the loss amount in the Merton model. Figure (a) is the overview. Figure (b) is the detail  of $f_i$. The first, second and fourth registers are $R_{\rm PRN},R_{\rm com}$ and $R_L$ respectively. The third register, to which the result of comparison is output, is omitted in (a).}
	\label{MertonCircuit}
\end{figure*}

\subsection{Merton Model}

Now, let us consider the application of the aforementioned method to the actual problem in finance.
We take credit risk measurement, which is mentioned in the introduction, as an example.
First, we briefly explain the Merton model\cite{Merton}, which is widely used in practice in many banks.

In this model, the stochastic loss amount $L$ in a credit portfolio consisting of $N_{\rm obl}$ obligors is given as follows:
\begin{eqnarray}
L & = & \sum_i^{N_{\rm obl}} E_i{\bf 1}_{Z_i<z_i}, \nonumber \\
Z_i & = & \alpha_i \epsilon_{\rm com} + \sqrt{1-\alpha_i^2} \epsilon_i. \label{loss}
\end{eqnarray}
The meaning of each symbol is as follows.
$E_i$ is the exposure of the $i$ th obligor, that is, the loss arising if he defaults\footnote{Here, we assume that loss given default is 1.}.
${\bf 1}_C$ is the indicator function, which is 1 if the condition $C$ is satisfied and 0 otherwise.
The stochastic variable $Z_i$ is interpreted as ``the value of the firm" for the $i$ th obligor.
We consider that he defaults if $Z_i$ becomes smaller than a threshold $z_i$.
Usually, given a probability of default $p_i$ exogenously, $z_i$ is set as $z_i=\Phi_{\rm SN}^{-1}(p_i)$, where $\Phi_{\rm SN}$ is the distribution function for standard normal distribution and $\Phi_{\rm SN}^{-1}$ is its inverse.
$Z_i$ is given as a linear combination of two independent standard normal random variables $\epsilon_{\rm com}$ and $\epsilon_i$.
$\epsilon_{\rm com}$ is common for all obligors and called a systematic risk factor, which is interpreted as a factor reflecting the situation of macro economy\footnote{Although we consider the case there is a single systematic risk factor, we can extend the model with multiple ones.}.
$\epsilon_i$ is called an idiosyncratic risk factor and represents the effect of the matters unique to the $i$ th obligor on his credit.
We take the coefficient $\alpha_i$ such that $0<\alpha_i<1$; therefore, $Z_i$ is also standard normal.
$\alpha_i$ determines the correlation between $Z_i$ for different obligors: the larger $\alpha_i$ means stronger correlation and a larger probability of simultaneous defaults of many obligors.

\subsection{Calculation of Loss Using PRNG on Quantum Circuit}

Then, we describe how to calculate credit risk measures using a PRNG on a quantum circuit.
What we have to develop is the circuit which calculates stochastic loss amount $L$.
Once we develop the circuit which creates a superposition of states in which the value of the loss is encoded in some register, we can estimate VaR and CVaR as explained in \cite{EggerEtAl}.
The difference between the way in this paper and those in previous works is how to create such a superposition.

Seeing (\ref{loss}), we notice that the loss $L$ can be written as a sum of contributions from each obligor and takes the form of (\ref{integrandForm}) as mentioned above.
More concretely, precomputing $\epsilon_{\rm com}$ and defining
\begin{eqnarray}
f_1(x) & = & E_1{\bf 1}_{x<Y_1(\epsilon_{\rm com})} \nonumber \\
f_i(y, x) & = & y + E_i{\bf 1}_{x<Y_i(\epsilon_{\rm com})}, i=2,...,N_{\rm obl}
\end{eqnarray}
we can calculate $L$ as
\begin{eqnarray}
y_1 & = & f_1(x_1) \nonumber \\
y_2 & = & f_2(y_1, x_2) \nonumber \\
& \vdots &  \nonumber \\
y_{N_{\rm obl}-1} & = & f_{N_{\rm obl}-1}(y_{N_{\rm obl}-2}, x_{N_{\rm obl}-1}) \nonumber \\
L & = & f_{N_{\rm obl}}(y_{N_{\rm obl}-1}, x_{N_{\rm obl}}),
\end{eqnarray}
using PRNs $x_1,...,x_{N_{\rm obl}}$ as $\epsilon_1,...,\epsilon_{N_{\rm obl}}$.
Here, $Y_i(\epsilon_{\rm com}) = M_{\rm PRN}P_i(\epsilon_{\rm com})$, where
\begin{equation}
P_i(\epsilon_{\rm com})=\Phi_{\rm SN}\left(\frac{z_i-\alpha_i\epsilon_{\rm com}}{\sqrt{1-\alpha_i^2}}\right).
\end{equation}
is the conditional probability that the $i$-th obligor defaults given $\epsilon_{\rm com}$ and $M_{\rm PRN}$ is the maximum number that the PRN can take.

The concrete calculation flow to obtain one sample value of $L$ is as follows:

\begin{enumerate}
	\item Generate a standard normal random variable and let it be $\epsilon_{\rm com}$.
	\item Set $i=1$ and $L=0$.
	\item Set the first elements of the PRN sequence $x_1$.
	\item Calculate $Y_i(\epsilon_{\rm com})$.
	\item Compare $x_i$ with $Y_i(\epsilon_{\rm com})$. If the former is smaller than the latter, update $L \leftarrow L+E_i$.
	\item If $i=N_{\rm obl}$, finish. Otherwise, progress the PRN sequence to get $x_{i+1}$, update $i \leftarrow i+1$ and go to 4.
\end{enumerate}

The above flow is performed by the circuit in Figure \ref{MertonCircuit}.
As explained in Section \ref{sec:overview}, we first create a superposition of $\ket{x^{(1)}_1}_{n_{\rm PRN}},...,\ket{x^{(N_{\rm samp}-1)}_1}_{n_{\rm PRN}}$ on $R_{\rm PRN}$ using $H^{\otimes n_{\rm{samp}}}$ and $J_{\rm PRN}$.
These are starting elements of PRN sequences.
Besides, we create a superposition of $x$, that is, numbers which obey the standard normal distribution in the register $R_x$.
This is done by the method described in \cite{Grover} and depicted as the gate 'SN' in Figure \ref{MertonCircuit}.
Then, progressing the PRN sequence by $P_{\rm PRN}$, $L$ is sequentially calculated by $f_1,...,f_{N_{\rm obl}}$.

In $f_i$, at first, $x$ is converted to $Y_i(x)$ by the gate $Y_i$\footnote{In practice, parameters such as $p_i$ and $\alpha_i$ are not set for each obligor individually. Instead, obligors are grouped in terms of industry sector or rating and the same parameter values are given in obligors in each group. In such a case, it is not necessary to operate $Y_i$ for each $i$, but sufficient to operate once per group.}.
We here simply assume that such a gate exists.
In \cite{Woerner, EggerEtAl}, several ways to calculate such a function on a quantum computer are proposed, for example, linear approximation or piecewise polynomial approximation \cite{Haner}.
Then $x_{i}$ is compared with $Y_i(x)$ and an ancillary qubit is set to 1 if $x_{i}>Y_i(x)$.
Such a comparator has been presented in \cite{Oliveira}.
With the control by the ancilla, $E_i$ is added to the loss register $R_L$.
The controlled adder is also presented in previous papers, such as \cite{Vedral}.
Finally, the inverses of $Y_i$ and comparison are performed to uncompute $R_x$ and the ancilla.

\section{\label{sec:demo}Demonstration: Application to Integration of Simple Multi-variable Function}

\begin{figure*}[t]
	\begin{minipage}{1\hsize}
		\begin{center}
\includegraphics[width=0.7\textwidth]{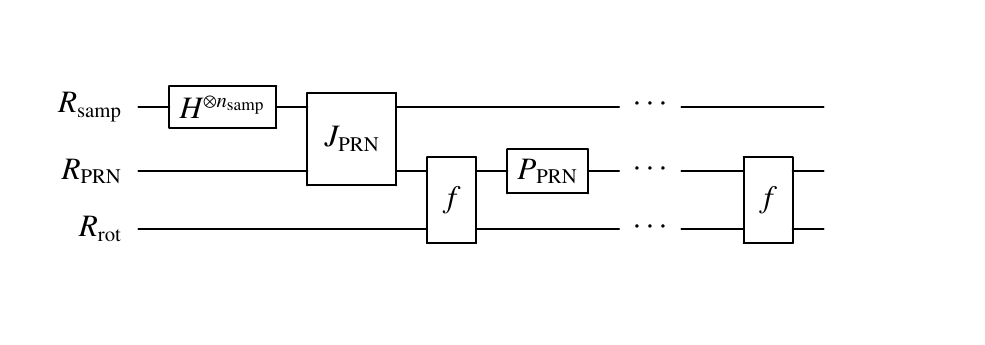}
		\end{center}
		\subcaption{}
	\end{minipage}\\	
	\includegraphics[width=1\textwidth]{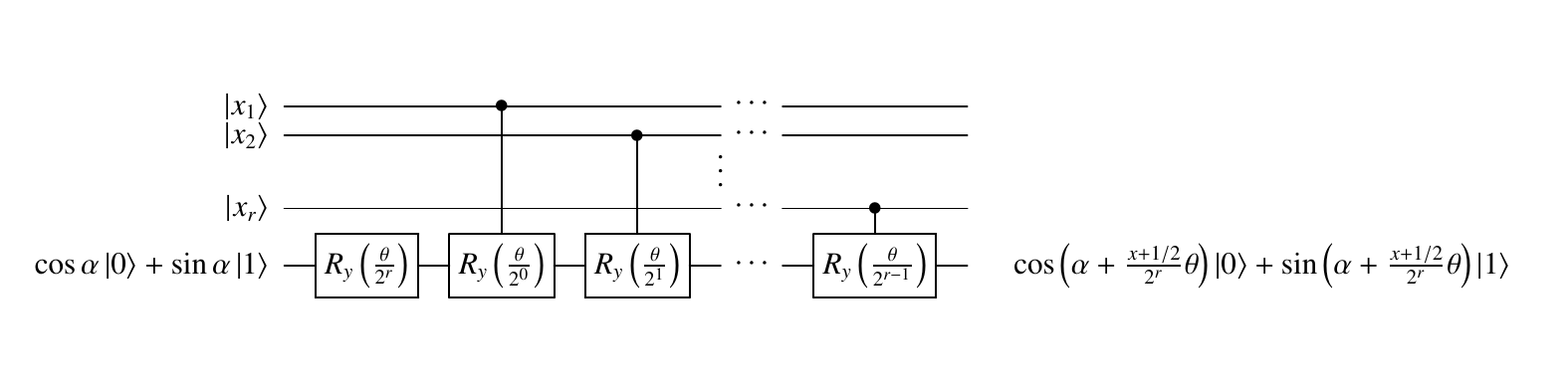}
	\begin{minipage}{1\hsize}
		\begin{center}
			
		\end{center}
		\subcaption{}
	\end{minipage}	
	\caption{The Quantum circuit for the integration (\ref{sinIntegMC}). Figure (a) is the overview. Figure (b) is the detail of $f$. Here, $\ket{x_1},...,\ket{x_r}$ are qubits in $R_{\rm PRN}$ used as a $r$ bit random number. $\ket{x_1}$ is the most significant digit and $\ket{x_r}$ is the least one.}
	\label{SinIntegCircuit}
\end{figure*}

\begin{table*}[t]
	\begin{tabular}{|l|r|} \hline
		(i) Exact value of the original integral & 0.074578 \\  \hline
		(ii) Exact average of integrand values on sample points & 0.078394 \\  \hline
		(iii) Estimate by the method in Section \ref{sec:demo} & 0.078391 \\ \hline
	\end{tabular}
	\caption{Values of the integral obtained in various ways.}
	\label{integResult}
\end{table*}

Although the method proposed in this paper reduces required qubits, the circuit presented in the last section is still too large to perform in simulators or machines which can be publicly used today.
We therefore consider a more small-scale problem performable in a simulator.
It is an integral of a trigonometric function
\begin{equation}
I=\frac{1}{\theta^{N_{\rm var}}}\int_0^{\theta}dx_1\cdots \int_0^{\theta}dx_{N_{\rm var}} \sin^2\left(\sum_{i=1}^{N_{\rm var}}{x_i}\right),
\label{originalI}
\end{equation}
which is the multi-variable version of the problem considered in \cite{Suzuki}.
Note that the phase in the sin function depends on $N_{\rm var}$ variables.
A naive way to calculate such a multi-variable integration numerically is discretization, that is, taking the sum of the integrand values on grid points set with equal interval in each axis,
\begin{equation}
I\simeq\frac{1}{N^{N_{\rm var}}}\sum_{i_1=0}^{N-1}\cdots \sum_{i_{N_{\rm var}}=0}^{N-1} \sin^2\left(\sum_{j=1}^{N_{\rm var}}{\frac{i_j+1/2}{N}}\theta\right),
\end{equation}
where $N$ is the number of intervals in each axis.

However, in a brute force summation like this, the computational load increases exponentially with the number of variables since the number of grid points is $N^{N_{\rm var}}$.
So the alternative way is Monte Carlo integration, that is, taking the average of the integrand values on grid points which are randomly sampled using PRNG.
More specifically, taking an $r$-bit PRN sequence $\{x_i\}_{i=1,2,...}$, where $x_i\in\{0,1,...,2^r-1\}$, we make the approximation as
\begin{equation}
I\simeq\frac{1}{N_{\rm samp}}\sum_{i=0}^{N_{\rm samp}-1} \sin^2\left(\sum_{j=1}^{N_{\rm var}}{\frac{x^{(i)}_j+1/2}{2^r}}\theta\right),
\label{sinIntegMC}
\end{equation}
where $x^{(i)}_j=x_{iN_{\rm var}+j}$.
Note that $(x^{(i)}_j+1/2)/2^r$ is the pseudorandom number which takes one of $2^r$ grid points in an axis, $1/2^{r+1},(1+1/2)/2^r,...,(2^r-1+1/2)/2^r$.

Note that we can use the method we proposed to calculate (\ref{sinIntegMC}), since (\ref{sinIntegMC}) is in the form of (\ref{integrandForm}).
Defining
\begin{eqnarray}
f_1(x) & = & \frac{x + 1/2}{2^r}\theta \nonumber \\
f_2(y,x) & = & y + \frac{x + 1/2}{2^r}\theta \nonumber \\
& \vdots & \nonumber \\
f_{N_{\rm var}-1}(y,x) & = & y + \frac{x + 1/2}{2^r}\theta \nonumber \\
f_{N_{\rm var}}(y,x) & = & \sin^2\left(y + \frac{x + 1/2}{2^r}\theta\right),
\end{eqnarray}
we can calculate a sample value of the integrand $f^{(i)}=\sin^2\left(\sum_{j=1}^{N_{\rm var}}{\frac{x^{(i)}_j+1/2}{2^r}\theta}\right)$ as 
\begin{eqnarray}
y^{(i)}_1 & = & f_1(x^{(i)}_1) \nonumber \\
y^{(i)}_2 & = & f_2(y^{(i)}_1, x^{(i)}_2) \nonumber \\
& \vdots &  \nonumber \\
y^{(i)}_{N_{\rm var}-1} & = & f_{N_{\rm var}-1}(y^{(i)}_{N_{\rm var}-2}, x^{(i)}_{N_{\rm var}-1}) \nonumber \\
f^{(i)} & = & f_{N_{\rm var}}(y^{(i)}_{N_{\rm var}-1}, x^{(i)}_{N_{\rm var}}). \label{flowSec5}
\end{eqnarray}

In fact, we perform the calculation in a slightly different way from (\ref{flowSec5}), since there is a more efficient way in this case.
The quantum circuit for the calculation is shown in Figure \ref{SinIntegCircuit}.
In the circuit, we do not compute the integrand value on a register then rotate the phase of an ancilla with control of the register as (\ref{overallflow}), but sequentially rotate the ancilla's phase according to the PRN value.
More concretely, the implementation is as follows. 
In addition to $R_{\rm samp}$ and $R_{\rm PRN}$, the circuit has an ancilla, which we hereafter write as $R_{\rm rot}$.
The value of the integration (\ref{sinIntegMC}) is encoded into its phase by the gate $f$ in Figure \ref{SinIntegCircuit}.
This gate is a sequence of rotations around $y$-axis $R_y$ controlled by output qubits in $R_{\rm PRN}$\footnote{Note that not all qubits in $R_{\rm PRN}$ represent output random numbers. For example, bottom bits in PCG are not used due to poor statistical property.}.
That is, if $R_{\rm rot}$ is in the state $\cos\alpha\ket{0} + \sin\alpha\ket{1}$ for some real number $\alpha$ and $R_{\rm PRN}$ is in the state corresponding to a random number $x$ before $f$, going through it transforms the state as follows:
\begin{eqnarray}
& & \cos\alpha\ket{0} + \sin\alpha\ket{1} \rightarrow \nonumber \\
& & \cos\left(\alpha + \frac{x+1/2}{2^r}\theta\right)\ket{0} + \sin\left(\alpha + \frac{x+1/2}{2^r}\theta\right)\ket{1},
\end{eqnarray}
that is, rotation by the angle $\frac{x+1/2}{2^r}\theta$.
Therefore, starting from the state in which all registers are 0, the entire circuit transforms the state as follows:
\begin{eqnarray}
& \ket{0}_{\rm all} \coloneqq \ket{0}_{n_{\rm samp}} &\ket{0}_{n_{\rm PRN}}\ket{0} \rightarrow \nonumber \\
& \frac{1}{\sqrt{N_{\rm samp}}}\sum_{i=1}^{N_{\rm samp}}&\ket{i}_{n_{\rm samp}}\ket{x^{(i)}_{N_{\rm var}}}_{n_{\rm PRN}} \nonumber \\
& & \otimes \Biggl[\cos\left(\sum_{j=1}^{N_{\rm var}}{\frac{x^{(i)}_j+1/2}{2^r}}\theta\right)\ket{0}  \nonumber \\ 
& & \qquad + \sin\left(\sum_{j=1}^{N_{\rm var}}{\frac{x^{(i)}_j+1/2}{2^r}}\theta\right)\ket{1}\Biggr].
\end{eqnarray}
So the probability to observe $\ket{1}$ is equal to (\ref{sinIntegMC}).

The probability to observe $\ket{1}$ can be estimated, for example, in the way proposed in \cite{Suzuki}, which we here explain briefly.
First we construct the operation
\begin{equation}
Q = -AS_0A^{-1}S_\chi,
\end{equation}
where $A$ corresponds to the entire circuit in Figure \ref{SinIntegCircuit}, $S_0$ multiplies $-1$ to the state if all qubits are 0 or does nothing otherwise and $S_\chi$ multiplies $-1$ to the state if $R_{\rm rot}$ is 1 or does nothing otherwise.
If we write the probability to observe $\ket{1}$ in $R_{\rm rot}$ in $A\ket{0}_{\rm all}$ as $\sin^2\theta_a$, where $\theta_a\in[0,\pi/2]$, that in $\ket{\Psi_m}\coloneqq Q^mA\ket{0}_{\rm all}$ is $\sin^2((2m+1)\theta_a)$. 
So, choosing a set of non-negative integers $m_0,m_1,...,m_M$ and making $N_k$ observations of $R_{\rm rot}$ in $\ket{\Psi_{m_k}}$ for each $m_k$, we can estimate $\theta_a$ as the maximum point of the following likelihood function:
\begin{equation}
L_{\rm lik}(\theta_a)\coloneqq\prod^M_{k=0}{\left[\sin^2((2m_k+1)\theta_a)\right]^{h_k}\left[\cos^2((2m_k+1)\theta_a)\right]^{N_k-h_k}},
\end{equation}
where $h_k$ is the number of observations where $R_{\rm rot}$ is $\ket{1}$ in $\ket{\Psi_{m_k}}$.

We have performed the actual calculation based on the above method using the quantum circuit simulator Qiskit of IBM \cite{Qiskit}.
The detailed setting is as follows.
We estimate the integral (\ref{originalI}) for $\theta=\pi/6$ and $N_{\rm var}=2$.
Although such a two-dimensional integral can be done analytically, the problem must be small-scale enough to be performed in the simulator and we consider it to be sufficient for a proof-of-concept.
For PRNG, we use LCG with parameters $a=11,c=0,m=31$ and the seed 1.
Then the PRN is 5-bit and the period is 30.
We take $N_{\rm samp}=8$ sample points, so using 16 elements in the PRN sequence.
Of course there are statistical concerns on the estimate based on such a small number of samples generated by such a small-scale PRNG, but it is inevitable in calculation on a simulator and sufficient for the current proof-of-concept purpose.
If we can use a real quantum computer with sufficient qubits, say 100, we should use PCG under an appropriate setting: with sufficiently many qubits (say, 32-bit output and 64-bit background LCG), widely-used LCG parameters and permutation recommended in \cite{PCG}.
For the implementation of LCG, we use the adder presented in \cite{Thapliyal} and construct modular adder, multiplier and exponentiator based on the adder following the way in \cite{Vedral}.
For $\theta_a$ estimation, we take $M=8,N_k=100$ and $m_k=2^k$, as in \cite{Suzuki}.

We show the result in Table \ref{integResult}.
At the time when the integral (\ref{originalI}) is approximated as (\ref{sinIntegMC}), some error arises.
This is the difference between (i) and (ii) in Table \ref{integResult}, which will become smaller if we can take more sample points generated by a larger-scale PRNG.
Estimation by quantum computer should converge to (ii), and the estimation obtained actually (iii) is close to (ii) as expected.

\section{\label{sec:ConclAndDisc}Conclusion and Discussion}

In this paper, we considered reduction of a qubit number in the quantum algorithm for Monte Carlo.
Although its applications to problems in finance are proposed in previous works, high-dimensionality of some of such problems requires many qubits if a quantum register is prepared for each of the random numbers required to calculate one sample value of the integrand.
Especially, for credit risk measurement, the required qubit number is proportional to the number of obligors, which can be $O(10^6)$.
Then we proposed a way to reduce the qubit number.
Considering the difference between what we calculate in the previous way of quantum-based Monte Carlo and that in classical Monte Carlo, we pointed out that estimating the average of sampled integrand values, which is calculated in classical Monte Carlo, by the quantum algorithm provides us with both quantum speed-up and qubit reduction.
We saw that such a way is realized by the PRNG on a quantum computer and presented a candidate for a PRNG implementable on a quantum computer, PCG, with concrete circuit diagrams.
We also described how to implement credit risk measurement using PRNG on quantum computer and demonstrated a simple integral on a quantum circuit simulator as a proof-of-concept.

As a final note, let us consider the trade-off between qubit number and circuit depth.
It is clear that qubit number reduction proposed in this paper increases circuit depth.
It is change of the design of the circuit, from that parallelly generate random numbers in different registers to that sequentially generate them in a register\footnote{Here, "parallel" means not parallel computation in quantum superposition but that in separate memories, which is possible also in classical computers.}.
Therefore, circuit depth is now proportional to the number of random numbers $N_{\rm ran}$\footnote{Note that, depending on problems, circuit depth can be proportional to $N_{\rm ran}$ even if random numbers are generated on different registers. That is, if there is no other way than calculating the integrand using random numbers in sequence, circuit depth is inevitably $O(N_{\rm ran})$, whether we generate random numbers sequentially or parallelly. Calculation of loss in a credit portfolio can be parallelized, as explained in \cite{EggerEtAl}.}.
This might make full reduction of qubit number by this way impractical.
Without quantum error correction \cite{Shor,Kitaev,Fowler}, which is expected not to be realized in near-term quantum computer, such a deep circuit will not be performable.
Even if a machine with error correction is developed, deep circuits might suffer from long runtime of fault-tolerant gates \cite{EggerEtAl, Fowler2} and quantum computation with small computational {\it load} might not necessarily lead to short computational {\it time}.

However, we consider the above trade-off itself meaningful.
Even if a quantum computer with large qubit number becomes in operation in the future, management of memory (that is, qubit) will be an important issue when it is applied to large-scale problems such as credit risk measurement.
That is, when fully parallel computation is impossible due to shortage of memory, we have to perform some procedures in sequence.
This is an issue which frequently arises also in today's classical computation.

The method proposed in this paper provides a way to solve such a problem in large-scale Monte Carlo simulation by quantum computer.
Consider the situation where $N_{\rm ran}$ random numbers are required to calculate the integrand but the available machine has so small a number of qubits that only $N_{\rm ran}/n$ random numbers can be generated at the same time, where $n$ is an integer satisfying $n\geq2$.
In such a case, we can parallelly generate $N_{\rm ran}/n$ PRN sequences with $n$ elements, calculate a part of the integrand using the elements in each sequence one-by-one, and finally merge partial results to obtain the entire integrand value\footnote{Again, this is possible only if the integrand allows such calculation.}.
This leads to the circuit depth proportional to $n$.
This is partial but maximum parallelism which can be done in the machine, although the depth is $n$ times larger than the full parallelism.

\section*{Acknowledgements}
The authors thank Shumpei Uno of Mizuho Information \& Research Institute and Kazuyoshi Yoshino, Naoyuki Takeda and Kazuya Kaneko of Mizuho-DL Financial Technology for helpful comments.

\end{document}